\documentclass[twocolumn,showpacs,amsmath,amssymb,prb,superscriptaddress]{revtex4}
\usepackage{graphicx}
\usepackage{bm}
\usepackage{dcolumn}
\usepackage{color}
\usepackage{braket}

\begin{document}
\title{Quantum mechanics of a spin-orbit coupled electron constrained to a space curve}

\author{Carmine Ortix}
\affiliation{Institute for Theoretical Solid State Physics, IFW-Dresden, Helmholtzstr. 20, D-01069 Dresden, Germany}

\begin{abstract}
We derive the effective one-dimensional Schr\"odinger-Pauli equation for electrons constrained to move on a space curve. The electrons are confined using a double thin-wall quantization procedure with adiabatic separation of fast and slow quantum degrees of freedom. This procedure is capable of yielding a correct  Hermitian one-dimensional Schr\"odinger-Pauli operator.
We find that the torsion of the space curve generates an additional quantum geometric potential, adding to the well-known curvature-induced one. 
Finally, we derive an analytic form of the one-dimensional Hamiltonian for spin-orbit coupled electrons in a nanoscale helical wire. 
\end{abstract}

\maketitle

\section{Introduction} 
 The experimental progress in synthesizing low-dimensional nanostructures with curved geometries \cite{sch01,ahn06,mei07,ko10,par10} has boosted the interest in the
quantum physics on curved low-dimensional spaces. Two theoretical approaches have been devised to describe the quantum mechanical properties 
of particles in curved $n$-dimensional spaces: a method due to De Witt \cite{dew57} that approaches the problem by studying the quantum dynamics as fully  $n$-dimensional, and another due to Jensen, Koppe \cite{jen71}, and  Da Costa \cite{dac82} (JKC) that treats the quantum motion as the limiting case of the motion in an ordinary $n+1$-dimensional Euclidean space subject to a strong confining force acting in the normal direction of a curved $n$-dimensional manifold. 
While the  De Witt-like approach leads to operator ordering ambiguities, the JKC formal description is well-defined and is obviously the most rigorous and physically sound one for curved two-dimensional (2D)  nanomaterials, which are embedded in the ordinary three-dimensional (3D) Euclidean space.

For non-relativistic electrons in curved 2D manifolds, the JKC thin-wall quantization procedure allows to derive an effective 2D Schr\"odinger equation where the effect of curvature is encoded in a quantum geometric potential (QGP), which causes intriguing phenomena at the nanoscale \cite{can00,aok01,enc03,fuj05,kos05,gra05,cha04,mar05,ved00,cuo09,ort10}.
In periodically minimal surfaces, for instance, the QGP leads to a topological band structure \cite{aok01}. Similarly, in spirally rolled-up nanotubes the QGP has been shown to lead to winding-generated bound states \cite{ort10}. These curvature effects have been predicted to become even more pervasive in strain-driven nanostructures where the nanoscale variation of strain induced by curvature leads to a strain-induced geometric potential that is of the same functional form as the QGP, but gigantically boosting it \cite{ort11b}. 

The JKC thin-wall approach has been recently shown to be well founded also in presence of externally applied electric and magnetic fields \cite{fer08,ort11} and subsequently employed to predict novel curvature-induced phenomena, such as the strongly anisotropic ballistic magnetoresistance of spirally rolled-up semiconducting nanotubes without magnetism and spin-orbit interaction \cite{cha14} . Finally, the experimental realization of an optical analog of the curvature-induced QGP has provided  empirical evidence for the validity of the JKC squeezing procedure \cite{sza10}. 

But in spite of its relevance to curved 2D nanostructures, the JKC formal description makes no assertion on the quantum mechanical properties of particles constrained to space curves,  a proper understanding of which has become immediate due to the present drive in constructing complex 3D nanoarchitectures, such as helical nanowires \cite{zha02} or multiple helices, toroids and conical spirals \cite{xu15}.  In this paper, we address this question and 
develop a JKC-like thin-wall quantization procedure for the electronic motion in a one-dimensional (1D) curved manifold embedded in the ordinary Euclidean 3D space. 
The electrons are confined to the space curve by the action of two strong confining potential in the normal and binormal directions. By subsequently employing a method of adiabatic separation of fast and slow quantum degrees of freedom, we show the appearance of a torsion-induced QGP which adds to the curvature-induced QGP. An immediate consequence of this result is the existence of different quantum mechanical properties for isometric planar and space curves with equal curvature profiles. The validity of our formal  procedure is 
demonstrated  by taking explicitly into account the electron spin-orbit coupling, and showing that our double thin-wall quantization method yields an Hermitian 1D Schr\"odinger-Pauli operator. 
Finally, we apply our procedure to derive the effective 1D Hamiltonian for electrons with spin-orbit coupling in a nanoscale helical wire \cite{qi09}.

\section{Torsion-induced quantum geometric potential}
In the usual effective-mass approximation  of semiconductors, the movement of electrons in presence of spin-orbit interaction can be described with an effective Schr\"odinger-Pauli equation acting on  a two-dimensional spinor $\psi$:
\begin{equation}
\left(\dfrac{{\bf p}^2}{2 m^{\star}} + \, {\boldsymbol \alpha} \cdot {\boldsymbol \sigma} \times {\bf p} \right) \psi = E \, \psi, 
\label{eq:ham0}
\end{equation}
where  ${\bf p} = -i \hbar \nabla$ is the canonical momentum operator and the $\boldsymbol{\sigma}$'s are the usual Pauli matrices generating the Clifford algebra of ${\mathcal R}^3$ , which obey the anticommutation relations $\left\{ \sigma_i, \sigma_j \right\} = 2 \, \eta_{i j}$ with  $\eta_{i j}$  the standard  spatial metric given by the identity matrix. In addition, we introduced  the vector ${\boldsymbol \alpha}$ whose direction and magnitude determine the spin-orbit field axis and spin-orbit interaction constant respectively.  
Finally $m^{\star}$ is the material dependent effective mass of the carriers. 
 In the remainder we will use Latin indices for spatial tensor components of the flat Euclidean three-dimensional space whereas Greek indices will be used for the corresponding tensor components in curved space. 
Adopting Einstein summation convention, Eq.~\ref{eq:ham0} can be generalized to a curved three-dimensional manifold as follows 
\begin{eqnarray}
E \psi&=& \left[ - \dfrac{\hbar^2}{ 2 m^{\star}} \left(G^{\mu \nu} \partial_\mu \partial_\nu - G^{\mu \nu} \,  \Gamma_{\mu \nu}^{\lambda} \partial_\lambda \right) \right.\nonumber \\ & & \left. -i \, \hbar \, \, {\mathcal E}^{\mu \nu \lambda}  \, \alpha_{\mu} \varsigma_{\nu} \partial_{\lambda} \right] \psi,
\label{eq:hamcurved}
\end{eqnarray}
where $G^{\mu \nu}$ is the inverse of the metric tensor $G_{\mu \nu}$, ${\mathcal E}^{\mu \nu \lambda}$ is the contravariant Levi-Civita tensor -- it can be written in terms of the usual Levi-Civita symbol as ${\mathcal E}^{\mu \nu \lambda} = \epsilon^{\mu \nu \lambda} / \sqrt{ || G || }$ -- and we introduced the affine connection
$$\Gamma_{\mu \nu}^{\lambda} = \dfrac{1}{2} G^{\lambda \xi} \left[\partial_\nu G_{\xi \mu} + \partial_{\mu} G_{\xi \nu} - \partial_{\xi} G_{\mu \nu} \right]. $$ 
Finally, the $\varsigma$'s are the generators of the Clifford algebra in curved space $\left\{\varsigma_\mu , \varsigma_{\nu} \right\} = 2 \, G_{\mu \nu}$.   

To proceed further, we need to define a coordinate system. We therefore start out by defining a space curve ${\mathcal C}$ of parametric equations ${\bf r}={\bf r}(s)$ with $s$ indicating the corresponding arclength. The portion of the three-dimensional space in the immediate neighborhood of ${\mathcal C}$ can be then parametrized as ${\bf R}(s ,q_2,q_3) = {\bf r}(s) +  \hat{N}(s) \, q_2 + \hat{B}(s) \, q_3,$ where $\hat{N}$ and ${\hat B}$ are the unit vectors normal and binormal to ${\cal C}$. The structure of the corresponding three-dimensional spatial metric tensor can be determined using that the tripod of orthonormal vectors ${\hat T}(s) = \partial_s {\bf r}(s)$, ${\hat N}(s)$ and ${\hat B}(s)$ obey a Frenet-Serret type equations of motion as they propagate along $s$
\begin{equation}
\left(
\begin{array}{c} \partial_s {\hat T}(s) \\ \partial_s {\hat N}(s) \\ \partial_s {\hat B}(s)  
\end{array}
\right) = \left( \begin{array}{ccc} 0 & \kappa(s) & 0 \\ 
-\kappa(s) & 0 & \tau(s) \\
0 & -\tau(s) & 0 
\end{array}
\right)
\left( \begin{array}{c}  {\hat T}(s) \\  {\hat N}(s) \\  {\hat B}(s) \end{array}  \right),
\end{equation}
where $\kappa(s)$ and $\tau(s)$ denote the curvature and torsion of the space curve respectively. With this, the metric tensor corresponding to the three-dimensional portion of space explicitly reads
\begin{equation*}
G=\left( \begin{array}{ccc} \left[1 - \kappa(s) q_2 \right]^2 + \tau(s)^2 \left(q_2^2+q_3^2\right) & -\tau(s) q_3 & \tau(s) q_2 \\
-\tau(s) q_3 & 1 & 0 \\ 
\tau(s) q_2 & 0 & 1 \end{array} \right),
\end{equation*}
whose determinant $||G||=\left[1 - \kappa(s) q_2 \right]^2$. The generators of the Clifford algebra for the metric tensor written above can be derived introducing the Cartan's dreibein formalism \cite{car04}.  At each point, we define  a set of one forms  with components $e_{\mu}^i$ and  a dual set of vector fields $e_i^{\mu}$ obeying the duality relations $e_\mu^i e^{\nu}_i = \delta^{\mu}_{\nu}$ and $e_{\mu}^i e_{j}^{\nu}= \delta_{i}^j$, and corresponding to the "square root" of the metric tensor $G_{\mu \nu} = e_{\mu}^{i} \delta_{i j } e_{\nu}^j$. The generators of the Clifford algebra can be then expressed as $\varsigma_\mu = e_\mu^i \sigma_i$. For the metric tensor written above, the dreibein field can be chosen as 
$e_{s}^{i} = \hat{T}^i(s) \left(1 - \kappa(s) q_2 \right) + q_2 \tau(s) \hat{B}^i(s) - q_3 \tau(s) \hat{N}^i(s)$,  $e_{q_2}^i= \hat{N}^i(s)$ and $e_{q_3}^i =  \hat{B}^i(s)$.  This immediately allows to identify the $\varsigma$'s  as $\varsigma_s = \sigma_T  \left(1 - \kappa(s) q_2 \right) + \sigma_B \tau(s) q_2  - \sigma_N \tau(s) q_3$, $\sigma_{q_2} = \sigma_N$, and $\sigma_{q_3}= \sigma_B $ written in terms of a local set of three Pauli matrices comoving  with the Frenet-Serret frame $\sigma_{T,N,B}= \boldsymbol{\sigma} \cdot (\hat{T}, \hat{N},\hat{B}$).  

In the same spirit of JKC \cite{jen71,dac82}, we now apply a thin-wall quantization procedure and take explicitly into account the effect of two strong confining potentials in the normal and binormal directions $V_{\lambda_N}(q_2)$, $V_{\lambda_B}(q_3)$ respectively, with $\lambda_{N,B}$ the two independent squeezing parameters. Furthermore, we introduce a rescaled spinorial wavefunction $\chi$ such that the line probability can be defined as $\int \chi^{\dagger} \chi \, d q_2 \, d q_3$. Conservation of the norm requires 
$${\cal N} = \int \sqrt{||G||} \, d s \, d q_2 \, d q_3 \, \psi^{\dagger} \psi =  \int d s \, d q_2\,  d q_3 \, \chi^{\dagger} \chi,$$
from which the rescaled spinor $\chi \equiv \psi \times ||G||^{1/4}$.

In the $\lambda_{N,B} \rightarrow \infty$ limit, the spinorial wavefunction will be localized in a narrow range close to $q_{2,3}=0$. This allows us to expand all terms appearing in Eq.~\ref{eq:hamcurved}  in powers of $q_{2,3}$. At the zeroth order we then obtain the following Schr\"odinger-Pauli equation: 
\begin{eqnarray}
E \, \chi&=& \left[ - \dfrac{\hbar^2}{ 2 m^{\star}} \left(\eta^{\mu \nu} \partial_\mu \partial_\nu + \dfrac{\kappa(s)^2}{4}  \right) -i  \hbar \, \, {\epsilon}^{\mu \nu \lambda}  \, \alpha_{\mu} \sigma_{\nu} \partial_{\lambda} \right.  \nonumber \\  & & \left.  -i \hbar \, \, {\epsilon}^{\mu \nu q_2}  \, \alpha_{\mu} \sigma_{\nu} \dfrac{\kappa(s)}{2} + V_{\lambda_N}(q_2) + V_{\lambda_B}(q_3)  \right] \chi
\label{eq:hamcurved2}
\end{eqnarray}
In the equation above, we have used that in the $q_{2,3} \rightarrow 0$ limit  the only non-vanishing affine connection component $\Gamma_{s \, s}^{q_2} = \kappa(s)$, and the limiting relations for the derivatives of the original spinor in terms of the rescaled one
\begin{equation*} 
\left\{ \begin{array}{l} \partial_{q_{2}} \psi = \partial_{q_2} \chi +\dfrac{\kappa(s)}{2} \chi  \\  \\
  \partial^2_{q_{2}} \psi = \partial^2_{q_2} \chi + \kappa(s) \partial_{q_2} \chi + \dfrac{3}{4} \kappa(s)^2 \chi.
\end{array} \right.
\end{equation*}
The presence of the relativistic spin-orbit interaction in Eq.~\ref{eq:hamcurved2} prevents  the separability of the quantum dynamics along the tangential direction of the space curve from the normal and binormal quantum motion. However, the strong size quantization along the latter directions still allows us to employ an adiabatic approximation \cite{ort11b}, encoded in the ansatz for the spinorial wavefunction $\chi(s,q_2,q_3)=\chi_T (s) \times \chi_N (q_2) \times \chi_B (q_3)$ where the normal and binormal wavefunctions solve the Schr\"odinger equation
$$-\frac{\hbar^2}{2 m^{\star}} \partial^2_{q_{2}, q_{3}} \, \, \chi_{N, B} + V_{\lambda_{N,B}} (q_{2,3}) \, \chi_{N,B} = E_{N,B} \, \chi_{N,B}.$$
We can assume the two confining potential to take either the form of an harmonic trap $\propto q_{2,3}^2$ or an infinite potential well centered at $q_{2,3} \equiv 0$. Taken perturbatively, the first derivatives terms $\partial_{q_{2,3}}$ of Eq.~\ref{eq:hamcurved2} vanish and thus the effective one-dimensional Schr\"odinger-Pauli equation for the tangential wavefunction reads
\begin{eqnarray}
E \, \chi_T &=&  \left[ - \dfrac{\hbar^2}{ 2 m^{\star}} \left(\partial^2_{s} + \dfrac{\kappa(s)^2}{4}  \right) -i \hbar \alpha_N \sigma_B \partial_s \right. \label{eq:hamcurved1D1} \\ & & \left. + i \hbar \alpha_T \sigma_B \dfrac{\kappa(s)}{2} + i \hbar \alpha_B \left( \sigma_N \partial_s - \sigma_T \dfrac{\kappa(s)}{2} \right) \right] \chi_T. 
\nonumber
\end{eqnarray}
For $\alpha_T \equiv 0$ and in the limit of zero torsion, {\it i.e.} for a planar curve,  Eq.~\ref{eq:hamcurved1D1} 
represents the correct effective one-dimensional Schr\"odinger-Pauli equation for a single electron in presence of spin-orbit interaction \cite{mei02,fru04,gen13} with the addition of the curvature-induced QGP. The corresponding Schr\"odinger-Pauli operator is indeed Hermitian as can be shown by calculating its matrix elements in any complete basis. 
However, as soon as the  curve fails to be planar or a finite $\alpha_T$ is taken into account, Hermiticity is lost --  the motion of electrons along a space curve cannot be described by Eq.~\ref{eq:hamcurved1D1}. 

We now show that this apparent paradox can be solved by expanding Eq.~\ref{eq:hamcurved} up to linear order in $q_{2,3}$.  
When averaged over the normal and binormal wave functions, these terms generally vanish. The exception are terms of the form $q_{2,3} \partial_{q_{2,3}}$ since they give rise to a finite contribution $\braket{q_{2,3} \partial_{q_{2,3}}} = -1/2$, {\it independent} of the specific form of the confining potential and its relative strength. Let us start with the spin-orbit interaction term. When expanded to linear order in $q_{2,3}$, we get the following correction, which we write in Hamiltonian form: 
\begin{eqnarray*}
\delta {\cal H}_{SO}& =&  - i \hbar \, \kappa(s) \, q_2 \, \epsilon^{\mu \nu \lambda} \alpha_\mu \sigma_\nu \partial_\lambda - i \hbar \epsilon^{\mu s \lambda} \alpha_\mu \partial_\lambda \times \\ & & \left[-\sigma_T \kappa(s) q_2 +  \sigma_B \tau(s) q_2 - \sigma_N \tau(s) q_3 \right], 
\end{eqnarray*}
where the first term originates from the covariant Levi-Civita tensor appearing in Eq.~\ref{eq:hamcurved}, while the second term is due to the structure of the Clifford algebra generator $\varsigma_s$ at $q_{2,3} \neq 0$. The only non-vanishing contributions of the equation above then read
\begin{equation}
\delta {\cal H}_{SO} =  - i \hbar  \alpha_T \sigma_B \dfrac{\kappa(s)}{2} + i \hbar \alpha_B \sigma_B \dfrac{\tau(s)}{2} + i \hbar \alpha_N \sigma_N \dfrac{\tau(s)}{2}. 
\label{eq:correctionSO}
\end{equation}

Next, we have to take into account the analogous corrections to the Laplace-Beltrami operator, which defines the kinetic energy of the electrons. In doing so, we first notice that up to linear order in $q_{2,3}$  the limiting relation for the second derivative of the original spinor in terms of the rescaled one has to be corrected as
$\partial^2_{q_{2}} \psi=  \partial^2_{q_2} \chi + \kappa(s) \partial_{q_2} \chi + \frac{3}{4} \kappa(s)^2 \chi  + \kappa(s)^2 q_2 \partial_{q_{2}}.$
It is then easy to show that the correction to the zeroth-order Laplace-Beltrami operator has the form 
\begin{eqnarray}
\delta {\cal H}_{LP} &=&- \dfrac{\hbar^2}{2 m^{\star}} \left[  \left( - 2 \tau(s)^2 q_2 \partial_{q_{2}} q_3 \partial_{q_{3}} + \kappa(s)^2 q_2 \partial_{q_{2}} \right)  \right. \nonumber \\
& &\left.+ \tau(s)^2 q_3 \partial_{q_{3}} + \left(\kappa(s)^2 + \tau(s)^2 \right) q_2 \partial_{q_{2}} \right], 
\label{eq:correctionKin}
\end{eqnarray}
where the first term in the r.h.s. originates from the non-diagonal components of the metric tensor $G^{\mu \nu}$ at $q_{2,3} \neq 0$, the second term follows from the aforementioned limiting relation for the second derivative of the spinor along the normal direction, while the third and fourth term are due to the affine connection components $\Gamma_{\mu \nu}^{q_{2}}$ and $\Gamma_{\mu \nu}^{q_{3}}$ respectively. Eq.~\ref{eq:hamcurved1D1},  Eq.~\ref{eq:correctionSO}, and Eq.~\ref{eq:correctionKin} averaged over the normal and binormal wavefunctions, define the effective one-dimensional Schr\"odinger-Pauli equation for the tangential wavefunction 
\begin{eqnarray}
E \, \chi_T &=&  \left[ - \dfrac{\hbar^2}{ 2 m^{\star}} \left(\partial^2_{s} + \dfrac{\kappa(s)^2}{4} + \dfrac{\tau(s)^2}{2}   \right) \right. \nonumber \\ & & \left. -i \hbar \alpha_N \left( \sigma_B \partial_s - \sigma_N \dfrac{\tau(s)}{2} \right) \right.  \label{eq:hamcurved2D1} \\ & & \left. + i \hbar \alpha_B \left( \sigma_N \partial_s - \sigma_T \dfrac{\kappa(s)}{2} + \sigma_B \dfrac{\tau(s)}{2}  \right) \right] \chi_T. 
\nonumber
\end{eqnarray} 
From the equation above, it is clear that the effect of terms linear in $q_{2,3}$ is twofold. First, the Schr\"odinger-Pauli operator in Eq.~\ref{eq:hamcurved2D1} is Hermitian for a generic space curve {\it independent} of the spin-orbit field axis. This can be shown by explicitly computing its matrix elements in any complete basis or simply noticing that it can be written in the compact form 
\begin{eqnarray*}
E \chi_T &=& \left[ \dfrac{\hat{p}_s^2}{2 m^{\star}} - \dfrac{\hbar^2 \left(\kappa(s)^2 +  2 \tau(s)^2 \right)}{8 m^{\star}} +\dfrac{\alpha_N}{2} \left\{\hat{p}_s , \sigma_B \right\} \right. \\ & & \left.  -\dfrac{\alpha_B}{2} \left\{{\hat p}_s , \sigma_N \right\} \right] \chi_T, 
\end{eqnarray*}
where we introduced the tangential momentum operator ${\hat p}_s=-i \hbar \partial_s$, and we have used the equation of motion for the set of three Pauli matrices comoving with the Frenet-Serret frame.
Second, Eq.~\ref{eq:hamcurved2D1} shows explicitly the appearance of a novel geometric potential that is quantum in nature -- it is proportional to $\hbar$ --  and is induced solely by the torsion of the space curve. This torsion-induced QGP appears in conjunction with the well-known curvature induced QGP, and implies intrinsically different quantum mechanical properties for electrons moving in two-dimensional and three-dimensional curves. 

\section{Helical nanowires}
Next, we apply the foregoing analysis to the paradigmatic case of a right-handed helical nanowire \cite{qi09} with parametric equation in cylindrical coordinates
\begin{equation}
\left\{ \begin{array}{l} x = R \,  \cos{\phi} \\ y = R \sin{\phi} \\ z = c \,  \phi \end{array} \right. ,
\end{equation}
with helix radius $R$ and  $2 \pi c$ pitch. The constant curvature $\kappa = R / (R^2 + c^2)$ while the torsion is given by $\tau = c / (R^2+c^2)$.  
Using that the arclength of the helix is related to the azimuthal angle $\phi$ by $s = \phi \,\,\sqrt{R^2+ c^2}$,  the Frenet-Serret frame is specified by the tripod of orthonormal vectors 
\begin{eqnarray*}
{\hat T}(\phi) &=& \left\{ -\cos{\alpha} \sin{\phi} , \cos{\alpha} \cos{\phi}, \sin{\alpha} \right\} \\ 
{\hat N}(\phi)&=& \left\{ -\cos{\phi} , \sin{\phi} , 0 \right\} \\ 
{\hat B}(\phi)&=& \left\{\sin{\alpha} \sin{\phi} , - \sin{\alpha} \cos{\phi} , \cos{\alpha} \right\},
\end{eqnarray*}
where we introduced the angle $\alpha = \arctan{(\tau / \kappa)}$. This allows us to write the effective 1D Hamiltonians for electrons with spin-orbit coupling in a nanohelix as
\begin{eqnarray}
{\cal H}_{SP} &=& -\, \Omega\,\left[ \partial^2_{\phi} +\dfrac{\cos{\alpha}^2}{4} + \dfrac{\sin{\alpha}^2}{2} \right] \nonumber \\ & &  - i \omega_N \left(\sigma_B \partial_\phi - \dfrac{\sin{\alpha}\, \sigma_N}{2} \right) \label{eq:hamiltonianhelix} \\ & &  + i \omega_B \left(\sigma_N \partial_\phi - \dfrac{ \sigma_T \, \cos{\alpha}}{2} + \dfrac{ \sigma_B \, \sin{\alpha}}{2} \right), \nonumber
\end{eqnarray}
where we defined $\Omega = \hbar^2 / [2 m^{\star} (R^2+ c^2)]$, and $\omega_{N,B} = \hbar \alpha_{N,B} / \sqrt{R^2 + c^2}$. For $\alpha \equiv 0$ the equation above corresponds precisely to the effective Hamiltonian in a one-dimensional quantum ring with intrinsic and curvature-induced Rashba spin-orbit interaction due to strain effects \cite{gen13}. The Hamiltonian Eq.~\ref{eq:hamiltonianhelix} can be solved using a trial spinorial wavefunction of the form $\chi = \left[ \chi_1 \mathrm{e}^{i \left(m - 1/2\right) \phi} , \chi_2 \mathrm{e}^{i \left( m + 1/2 \right) \phi} \right]$, where $m$ can assume only half-integer values in order to fulfill periodic boundary conditions, and the amplitudes $\chi_{1,2}$ depend explicitly on $m$. The corresponding energy spectrum can be simply found as 
\begin{eqnarray} 
E_{\pm} (m) &=& \Omega \left(m^2 - \dfrac{\sin{\alpha}^2}{4}  \right) - \dfrac{\omega_N}{2} \cos{\alpha} \pm |m| \times \nonumber \\ 
& & \sqrt{\Omega^2 + \omega_N^2 + \omega_B^2 - 2 \omega_N \Omega \cos{\alpha}}, \label{eigenvalueshelix}
\end{eqnarray}
which explicitly shows that the chemical potential and the electron spin-orbit splitting can be geometrically controlled with the  torsion of the nanohelix. 
 
\section{Conclusions}
We have derived, in conclusion, the effective one-dimensional Schr\"odinger-Pauli equation for electrons  constrained to move along a space curve using a double thin-wall quantization procedure and adiabatic separation of fast and slow quantum degrees of freedom. 
We have shown that the torsion of a space curve leads to an attractive quantum geometric potential, adding to the well-known curvature-induced geometric potential. As a result, the quantum mechanical properties of electrons confined to three-dimensional curves and two-dimensional curves can be intrinsically different. 
The validity of our formal procedure has been demonstrated by showing that it predicts the correct Hermitian Schr\"odinger-Pauli operator. Therefore, our method can be applied without restrictions for instance to study the electronic and transport properties of the recently synthesized three-dimensional complex nanoarchitectures of Ref.~\onlinecite{xu15}.

\section{Acknowledgements} 
I thank Jeroen van den Brink for stimulating and fruitful discussions, and acknowledge the financial support of the Future and Emerging Technologies (FET) programme within the Seventh Framework Programme for Research of the European Commission, under FET-Open grant number: 618083 (CNTQC).

\end{document}